\documentstyle[11pt,newpasp,twoside,epsf]{article}
\begin{document}
\title{Magnetic field decay, or just period--dependent beaming?}
 \author{Joeri van Leeuwen and Frank Verbunt}
\affil{Astronomical Institute, Utrecht University, PO Box 80000,
  3508 TA Utrecht, The Netherlands}

\begin{abstract}
   Several recent papers conclude that radio-pulsar magnetic fields
decay on a time-scale of 10\,Myr, apparently contradicting earlier
results. We have implemented the methods of these papers in our code
and show that this preference for rapid field decay is caused by the
assumption that the beaming fraction does not depend on the
period. When we do include this dependence, we find that the observed
pulsar properties are reproduced best when the modelled field does not
decay.

  When we assume that magnetic fields of new-born neutron stars are from
a distribution sufficiently wide to explain magnetars, the magnetic
field and period distributions we predict for radio are pulsars wider
than observed. Finally we find that the observed velocities
overestimate the intrinsic velocity distribution.
\end{abstract}

\section{Introduction}

\index{pulsars}
\index{magnetic fields}
\index{pulsars!magnetic fields}
\index{neutron stars!magnetic fields}
\index{anomalous X-ray pulsars!magnetic fields}
\index{radio}
\index{X-rays}
\index{pulsars!statistics}
\index{pulsars!surveys}
\index{pulsars!ages}
\index{pulsars!velocities} 

Neutron-star magnetic fields are determined in both x-ray and radio
pulsars. The cyclotron resonance scattering features in the 
x-ray spectrum of accreting neutron stars allow an estimate of the field
strengths. The results lie in a narrow range between (1-4)$ \times
10^{12}$\,G (Makishima et~al.\ 1999), for both young systems and systems as old as
$10^8$ years. This strongly suggests that neutron-star magnetic fields
do not decay spontaneously.

In contrast, three arguments have been put forward for rapid ($\sim
10$\,Myr) field decay in radio pulsars. Firstly, the anti-correlation
of characteristic age ($\tau_c \scriptstyle \propto P \dot
P^{\scriptscriptstyle -1}$) and magnetic field strength, or more
correctly, torque ($\scriptstyle \propto (P \dot P)^{\scriptscriptstyle
1/2}$), seems to indicate magnetic field decay. It is better explained
by the strong dependence of the quantities plotted.

Secondly, field decay could explain the scarcity of far-away pulsars,
as they stop shining before they cover large distances. Yet since
these distances are derived from dispersion measures, they will be
systematicly underestimated for pulsars beyond the galactic gas layer
(Bhattacharya \& Verbunt 1991), causing an apparent lack of far-away pulsars.

Thirdly, if magnetic fields or torques do not decay, one expects more
pulsars with long periods than are observed. A decrease in visibility
with time could also explain this shortage; in fact, both Vivekanand
\& Narayan (1981) and Lyne \& Manchester (1988) find that pulsars with
longer periods have significantly narrower beams, making it less
likely for them to be detected.

\section{The population synthesis code}

In a pulsar-population synthesis code, one models the birth, life and
death of radio pulsars and compares each simulated sample with the real
sample. Certain pulsars (nearby, slow, bright ones for example) are
more easily discovered than others and will be overrepresented in the
sample. To account for these selection effects, one must explicitly
model the surveys in which pulsars are detected.

Our code is a parallelised version of the one described by
Bhattacharya et~al.\ (1992) and Hartman et~al.\ (1997) and runs on
TERAS, the 1024-processor 1 Tflop/s supercomputer at SARA. In short,
the algorithm is as follows: in the initial settings we assume
distributions for the pulsar properties at birth, and fix the
time-scale for exponential magnetic field decay. The properties of
each simulated pulsar are drawn from the initial distributions.  We
evolve the magnetic field and period, calculate the orbit though the
galaxy and determine the size of the radio beam. We then check whether
any of the incorporated surveys discover the pulsar, and repeat the
loop until we have detected 2000 simulated pulsars. We compare their
properties to those of real pulsars, vary the input settings and
restart the synthesis, until the optimum solution is reached.

\section{Magnetic field decay}

Bhattacharya et~al.\ (1992) and Hartman et~al.\ (1997) found that models with long
magnetic field decay times reproduced the observations best. Forcing the use of
a short decay time, the simulation would only resemble the observed
data after the introduction of a second pulsar population, with longer
periods at birth, as was also found by Narayan \& Ostriker (1990).


Cordes \& Chernoff (1998) investigate the velocities and spin-down laws in 49 young
pulsars. They assume a constant beaming fraction and find a magnetic
field decay time of less than 10 Myr. Arzoumanian et~al.\ (2002) use a
period-dependent beaming model to generate pulse profiles. They do
not, however, use this new beaming model to fit for the magnetic field
decay time, but mention the value found by Cordes \& Chernoff (1998) instead:
less than 10 Myr. We have implemented pulse profiles in
the way described by Arzoumanian et~al.\ (2002), but still find that the observed
pulsar properties are best reproduced in a model without field
decay. However, if we follow Cordes \& Chernoff (1998) in assuming that the
beaming fraction does not depend on the period, we reproduce their
result and find that the best model has a decay time of order 10\,Myr.

 Arzoumanian et~al.\ (2002) further argue that all pulsars with the same
$\scriptstyle P$ and $\scriptstyle \dot P$ have the same intrinsic
(angle-averaged) luminosity and that the observed luminosity depends
on the observation angle and interstellar scintillation. As the width
of the resulting observed luminosity distribution is of the same
magnitude as the intrinsic spread in the Narayan \& Ostriker (1990) luminosity
function, the two descriptions are computationally almost identical.

Gonthier et~al.\ (2002) simulate galactic populations of radio and gamma-ray
pulsars. With a beaming fraction constant with period, their best
model has magnetic field decay in 5\,Myr.

Although the exact relation is still subject of debate, it is clear
that the beaming fraction decreases with the pulse period (Vivekanand
\& Narayan 1981, Lyne \& Manchester 1988). If we force our code to
disregard this relation, as Cordes \& Chernoff (1998), Gonthier
et~al.\ (2002) and Arzoumanian et~al.\ (2002) chose to do, we also
need magnetic-field decay on a time-scale of around $10^7$yr to
reproduce the observed period distributions.

When we do include period dependence of the beaming fraction and
search for the best solution, the no-decay case is much more probable
than decay on a 10\,Myr time scale (van Leeuwen \& Verbunt 2004).

\section{Magnetars}

	\begin{figure}[tb]
        \plotone{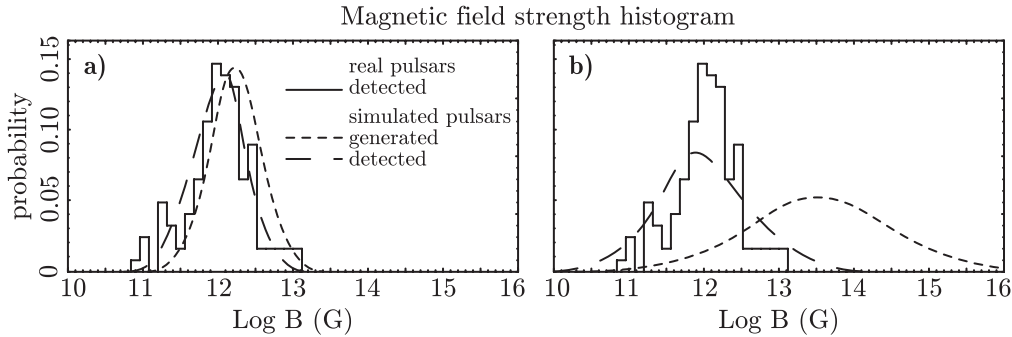}
        \caption{Single magnetic field distributions for the best
	model for {\bf a)} radio pulsars (Kolmogorov-Smirnov
	probability $p=0.56$) {\bf b)} radio pulsars and magnetars
	($p=0.03$)
        \label{img:BPhist}
	}
        \end{figure}

Most of the pulsars we detect in our code were born near the sun,
which makes us mostly sensitive to the local birthrate. In our best
model (no magnetic field decay) we need a local birthrate of 2
kpc$^{-2}$ Myr$^{-1}$, which can be produced by OB-association stars
with masses over $10\,M_{\odot}$ (Hartman et~al.\ 1997). In this
model, the magnetic field strength is drawn from a log-normal
distribution with centre and width $10^{12.3 \pm 0.3}$\,G. This makes
no neutron stars with magnetic fields higher than $10^{14}$\,G. To
make a substantial number of these, as well as normal radio pulsars,
we can shift and widen the distribution to $10^{13.5 \pm
0.9}$\,G. This increases the underlying neutron-star birthrate by
roughly a factor of 4. However, as the input magnetic field
distribution is wider, the simulated period distribution and magnetic
field distribution are both unacceptably wider than is observed (see
Fig. \ref{img:BPhist}). It therefore seems implausible that radio
pulsars and magnetars are formed from a single neutron-star magnetic
field distribution.

\section{Velocities}

\begin{figure}[tb]
        \plotone{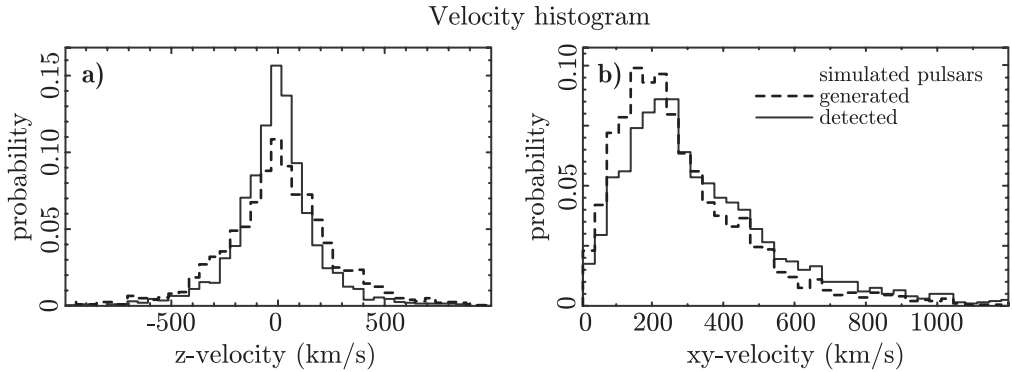}
        \caption{Velocity distribution in directions {\bf a)} perpendicular and {\bf b)}
	parallel to the galactic plane, for the generated and detected populations of
	simulated radio pulsars.
                     \label{img:velo}
	    }
        \end{figure}

In previous work on pulsar speeds (Helfand \& Tademaru 1977, Lyne \&
Lorimer 1994, Cordes \& Chernoff 1998) one argument keeps returning:
fast pulsars quickly move away from the plane, and so from our
telescopes, either to escape our galaxy or to spend much time far away
from it while they reverse direction. Many of these far-away pulsars
escape detection, biasing the observed sample to the slow and
sedentary pulsars (see Fig. \ref{img:velo}a)

This argument is valid in the z-direction only; in a three-dimensional
world, the situation is different (see also Hansen \& Phinney 1997). Seen
from the earth, most pulsars are born towards the galactic centre. Of
these, the high-speed ones can travel the largest distances and these
diffuse from the centre outward, towards our part of the galaxy
(Fig. \ref{img:velo}b). We find the effect of radial drift to be
stronger than the observational selection effect in the z-direction:
on the whole, our simulated detected pulsars move faster
than the parent population (van Leeuwen \& Verbunt 2004).

\end{document}